\documentclass[twocolumn,showpacs,nofootinbib,amsmath,amssymb,aps,prd,superscriptaddress,balancelastpage]{revtex4}

\usepackage{graphicx}
\usepackage[caption=false]{subfig}
\usepackage{amsmath,amssymb}
\usepackage{amsfonts,amssymb,mathrsfs}
\usepackage[bookmarks=false,pdfstartview=FitH]{hyperref}

\def\be{\begin{equation}}
\def\ee{\end{equation}}

\def\f{\frac}
\def\tf{\tfrac}

\def\d{\dot}

\def\t{\tilde}

\def\dd{{\rm d}}

\def\om{\omega}

\def\mH{\mathcal{H}}

\usepackage{color}

\begin{document}

\pagestyle{plain}

\title{Non-Singular Bounce Scenarios in Loop Quantum Cosmology\\ and the Effective Field Description}

\author{Yi-Fu Cai} \email{yifucai@physics.mcgill.ca}
\affiliation{Department of Physics, McGill University, Montr\'eal, QC H3A 2T8, Canada}

\author{Edward Wilson-Ewing} \email{wilson-ewing@phys.lsu.edu}
\affiliation{Department of Physics and Astronomy, Louisiana State University, Baton Rouge, 70803 USA}

\begin{abstract}
A non-singular bouncing cosmology is generically obtained in loop quantum cosmology due to non-perturbative quantum gravity effects. A similar picture can be achieved in standard general relativity in the presence of a scalar field with a non-standard kinetic term such that at high energy densities the field evolves into a ghost condensate and causes a non-singular bounce. During the bouncing phase, the perturbations can be stabilized by introducing a Horndeski operator. Taking the matter content to be a dust field and an ekpyrotic scalar field, we compare the dynamics in loop quantum cosmology and in a non-singular bouncing effective field model with a non-standard kinetic term at both the background and perturbative levels. We find that these two settings share many important properties, including the result that they both generate scale-invariant scalar perturbations. This shows that some quantum gravity effects of the very early universe may be mimicked by effective field models.
\end{abstract}


\maketitle

\section{Introduction}
\label{s.intro}

High precision observations of the cosmic microwave background (CMB) \cite{Spergel:2003cb, Ade:2013zuv} and the large scale structure provide strong evidence that the structure of our universe was seeded by a nearly scale-invariant power spectrum of primordial curvature perturbations. Such a primordial power spectrum can be causally generated from quantum vacuum fluctuations during an era of cosmic inflation \cite{Mukhanov:1981xt} where it is suggested that our universe experienced an exponentially accelerating expansion right after the initial big-bang singularity \cite{Guth:1980zm}. Alternatively, as suggested in \cite{Wands:1998yp, Finelli:2001sr}, a scale-invariant power spectrum of primordial perturbations can also be achieved during a matter-dominated contracting epoch, with the current expanding universe following a bouncing phase. A model of this type with an initial matter-dominated contraction and a non-singular bounce is called the matter bounce scenario, and provides an alternative to inflation for generating the observed spectrum of primordial fluctuations (see \cite{RHBrevs} for review articles on this topic).

The bouncing phase that connects the contracting and expanding branches can in principle be either singular (as in the original ekpyrotic scenario \cite{Khoury:2001wf}) or non-singular. Non-singular bouncing cosmologies appear in several settings where either the gravitational sector is modified as in Ho\u{r}ava gravity \cite{Calcagni:2009ar, Brandenberger:2009yt}, torsion gravity \cite{Cai:2011tc, Poplawski:2011jz} and non-local gravity \cite{Biswas:2005qr}, or the matter field violates positive energy conditions like in the quintom bounce \cite{Cai:2007qw, Cai:2007zv}, the ghost condensate bounce \cite{Buchbinder:2007ad, Lin:2010pf} and the Galileon bounce \cite{Qiu:2011cy} models.  For a recent review of non-singular bouncing cosmologies, see \cite{Novello:2008ra}.

Non-singular bounces also generically appear in loop quantum cosmology (LQC) \cite{Singh:2009mz}, where the variables and quantization techniques of loop quantum gravity are used to study quantum gravity effects in cosmological space-times \cite{Bojowald:2006da, Ashtekar:2011ni}. In LQC, the classical big-bang singularity  is replaced by a quantum bounce that occurs when the space-time curvature reaches the Planck scale \cite{Ashtekar:2006wn}. Moreover, the full quantum dynamics of semi-classical states are approximated to a high degree of accuracy by a simple set of effective equations \cite{Ashtekar:2006wn, Taveras:2008ke, Rovelli:2013zaa}. LQC is compatible with many interesting cosmological scenarios, including inflation \cite{Ashtekar:2009mm}, the matter bounce \cite{WilsonEwing:2012pu}, as well as the ekpyrotic scenario \cite{Wilson-Ewing:2013bla}, and therefore it is interesting to compare the predictions coming from the effective field models described above and from LQC.  As we shall see, many of the predictions are very similar.

An important conceptual issue for bouncing cosmologies is the well-known Belinsky-Khalatnikov-Lifshitz (BKL) anisotropic instability \cite{BKL}. The BKL instability appears in contracting cosmologies as the effective energy density contributed by the back-reaction of anisotropies increases faster than the energy densities of the dust and radiation matter fields. Therefore, in order to have a bounce that is approximately isotropic, it is necessary to fine-tune the initial conditions to be nearly perfectly isotropic in order to ensure that anisotropies never dominate. However, this problem is avoided in the ekpyrotic scenario where a scalar field with a steep and negative-valued potential always dominates over anisotropies in a contracting universe \cite{Erickson:2003zm} and so it is justified to neglect anisotropies in the presence of an ekpyrotic scalar field.

Recently, it has been shown how it is possible to combine an era of ekpyrotic contraction with a non-singular bounce by introducing a scalar field with a Horndeski-type non-standard kinetic term and a negative exponential potential \cite{Cai:2012va}. Furthermore, one may include a regular dust field and assume the universe began in a state of matter-dominated contraction thus combining the matter bounce with the ekpyrotic scenario. It can be explicitly checked that anisotropies remain small throughout the entire cosmological evolution of the matter-ekpyrotic bounce model, including at the bounce point \cite{Cai:2013vm}, and therefore this model successfully avoids the BKL instability that arises for a large family of non-singular bounce models, as pointed out in \cite{Xue:2010ux}. Among many possible implementations of the matter bounce, a concrete realization of the matter-ekpyrotic bounce was constructed in \cite{Cai:2013kja} that involves two matter fields with one being the scalar field that causes the bounce and the other representing the matter field that is dominant at the beginning of the contracting phase. Note that this effective field theory model of a non-singular bounce can also be developed into a super-symmetric version \cite{Koehn:2013upa}.

In this paper, we shall study the matter-ekpyrotic bounce within the context of LQC and compare its dynamics to those found in the effective field approach developed in \cite{Cai:2012va}. We find that the evolution of both the background and perturbative degrees of freedom are comparable in these two settings, and that in particular the dynamics of the primordial cosmological fluctuations in these two frameworks possess very similar features. For example, the sound speed squared, which characterizes the propagations of these fluctuation modes, becomes negative during the bouncing phase in both settings. This indicates that what appeared to be an instability in the effective field theory may in fact mimic a quantum gravity effect. In addition, the scale-dependence of the power spectrum studied in the effective field approach coincides with the one obtained in LQC. These similarities indicate that some of the quantum gravity effects of the very early universe can indeed be reproduced by effective field theory models.

In addition, since the ekpyrotic scalar field dominates the dynamics as the bounce is approached, the BKL instability is avoided as the back-reaction due to anisotropies will remain negligible compared to the energy density of the ekpyrotic scalar field. Because of this property of the matter-ekpyrotic bounce model, it is possible to neglect anisotropies and take the background to be the simplest homogeneous cosmology, namely the isotropic flat Friedmann-Lema\^itre-Robertson-Walker (FLRW) universe where the dynamics, of both the background cosmology and the perturbations, are significantly simpler.

The paper is organized as follows. In Sec.\ \ref{s.eff} we briefly review the matter-ekpyrotic bounce effective field theory model with the scalar field with a non-standard Horndeski-type kinetic term. Then in Sec.\ \ref{s.hom} we study the evolution of the matter-ekpyrotic homogeneous cosmology in LQC where the scalar field now has the standard kinetic term (and the potential is taken to be the same). We show explicitly how a non-singular bounce occurs near the Planck regime where quantum gravity effects become important. Afterwards, in Sec.\ \ref{s.scal} we analyze the dynamics of cosmological perturbations both analytically and numerically in the LQC matter-ekpyrotic model by tracking each Fourier mode throughout the cosmic evolution and show how the modes that exit the Hubble radius during the initial period of matter-dominated contraction are scale-independent. We close with some concluding remarks in Sec.\ \ref{s.disc}.

In this paper, we use units where $c = \hbar = 1$ but leave the reduced Planck mass $M_{\rm Pl}  = 1/\sqrt{8\pi G_N}$ (with $G_N$ being Newton's gravitational constant) explicit.  Also, the normalization of the scale factor is chosen so that its value at the bounce time $t=0$ is $a_{B} \equiv a(t=0) = 1$.


\section{The Effective Field Model of the Matter-Ekpyrotic Bounce}
\label{s.eff}

In this section, we briefly review how it is possible to obtain a non-singular bounce and avoid the BKL instability when the dominant matter field is a Horndeski-type scalar field with an ekpyrotic potential.  For more details concerning this model, see \cite{Cai:2012va}. We phenomenologically consider a scalar field with a Lagrangian of the type
\be \label{L_KGB}
 {\cal L} = K(\phi, X) + G(X)\Box\phi~,
\ee
and specifically choose the forms of the operators $K$ and $G$ to be
\begin{align} \label{KG}
 K(\phi, X) &= [1-g(\phi)] X +\f{\beta X^2}{M_{\rm Pl}^4} -V(\phi)~, \\
 G(X) &= \f{\gamma X}{M_{\rm Pl}^3} ~,
\end{align}
with $X$ being defined as the regular kinetic term $X \equiv g^{\mu\nu} (\partial_\mu\phi) (\partial_\nu\phi) /2$, while $\beta$ and $\gamma$ are coupling constants and $\Box \equiv g^{\mu\nu}\nabla_\mu\nabla_\nu$ is the standard d'Alembertian operator.

Note that in this model the $K$ operator involves the term $\beta X^2$ which can always stabilize the kinetic energy of the scalar field at high energy scales when $\beta$ is positive-definite. A bouncing phase can be triggered by allowing $g(\phi)$ to evolve for a short time into the regime of $g(\phi)>1$ which leads to the emergence of a ghost condensate. When this occurs, the null energy condition is violated and this can cause a non-singular bounce. One can design the form of $g(\phi)$ to be small far away from the bouncing phase so that the kinetic term in the Lagrangian of $\phi$ is well approximated by the standard kinetic term before and after the bounce. Additionally, the potential $V(\phi)$ plays the role of governing the dynamics of $\phi$ away from the bounce as well as determining the energy scale the bounce occurs at. In order to dilute the unwanted anisotropy to avoid the BKL instability, one takes the potential to have the ekpyrotic form of a negative exponential (at least for $\phi \ll -M_{\rm Pl}$).

To be specific, following \cite{Cai:2012va} we take the function $g(\phi)$ and the potential $V(\phi)$ to be
\begin{align}
 g(\phi) &= \frac{2g_0}{e^{-\sqrt{\frac{2}{q}}\frac{\phi}{M_{\rm Pl}}} +e^{b_g\sqrt{\frac{2}{q}}\frac{\phi}{M_{\rm Pl}}}}~, \\
 \label{gV}
 V(\phi) &= -\frac{2V_0}{e^{-\sqrt{\frac{2}{p}}\frac{\phi}{M_{\rm Pl}}} +e^{b_V\sqrt{\frac{2}{p}}\frac{\phi}{M_{\rm Pl}}}}~,
\end{align}
where $p, q, b_g, b_V$ and $g_0\equiv g(0)$ are dimensionless positive constants and $V_0$ is also positive with dimensions of $({\rm mass})^4$. We choose $g_0$ to be slightly bigger than unity so that the scalar can form a ghost condensate state when $\phi$ nears zero. The critical value of $g$ that signals the onset of the non-singular bouncing phase is therefore $g(\phi_*)=1$. By solving this equation, one finds the approximate values of $\phi_*$ where the non-singular bouncing phase begins and ends, respectively $\phi_{*-} \simeq -M_{\rm Pl}\ln (2g_0 / q) $ and $\phi_{*+} \simeq M_{\rm Pl}\ln (2g_0 /b_g q)$.  In addition, in order to obtain ekpyrotic contraction, we take $p \ll 1$.

Along with the ghost condensate phase, there are two interesting phenomena that occur around the bounce point. First, the proper (or cosmic) time derivative of the scalar field $\dot\phi$ reaches its maximal value around the bounce time $t_B$, as determined by the bounce condition that the Hubble rate $H$ vanishes at the bounce time, $H_B \equiv H(t_B)=0$. Second, the square of the sound speed decreases for a time to a negative value, approximately of the form
\begin{eqnarray}\label{csB}
 c_s^2(t_B) \simeq \frac{1}{3} - \frac{2}{3\sqrt{1+\frac{12\beta V_0}{M_{\rm Pl}^4(g_0-1)^2}}}~.
\end{eqnarray}
Thus, for a large class of parameter choices, $c_s^2(t_B)\simeq -1/3$. This may at first appear problematic as a negative value of the sound speed squared causes a dangerous exponential growth in the amplitude of short wavelength perturbations. However, if the bouncing phase lasts for a very short period, then this growth only lasts for a small time and so remains under control. As we shall see below, this phenomenon also occurs around the bounce in LQC as a result of quantum gravity effects.

\section{Homogeneous Loop Quantum Cosmology}
\label{s.hom}

In LQC, cosmological space-times are quantized by taking the holonomies of the $SU(2)$ Ashtekar-Barbero connection and areas to be the fundamental geometrical operators in the quantum theory. From the Hamiltonian constraint operator, one can derive the effective equations of motion that include the leading order quantum gravity corrections to the classical equations of general relativity \cite{Ashtekar:2011ni}.

For semi-classical states, namely states that are sharply peaked at late times when quantum gravity effects are negligible, the quantum dynamics at all times ---including at the bounce point where quantum gravity effects are strongest \cite{Rovelli:2013zaa}--- are very well approximated by a set of effective equations.  For a flat FLRW universe, the effective Friedmann equations that capture the salient quantum gravity effects of LQC are \cite{Ashtekar:2006wn, Taveras:2008ke}
\begin{align}
 H^2 &= \frac{1}{3M_{\rm Pl}^2} \rho \left( 1 - \f{\rho}{\rho_c} \right)~, \\
 \dot{H} &= - \frac{1}{2M_{\rm Pl}^2} (\rho + P) \left( 1 - \f{2 \rho}{\rho_c} \right)~,
\end{align}
while the continuity equation for the matter fields remains unchanged,
\begin{equation}
 \dot\rho + 3H(\rho+P) = 0~.
\end{equation}
Here $H = \d a / a$ is the Hubble rate and the dot denotes the derivative with respect to the proper time $t$. The critical energy density $\rho_c \sim M_{\rm Pl}^4$ is the maximum total energy density possible, and this upper bound is reached precisely at the bounce point where $\rho = \rho_c$, $H=0$ and $\dot{H} > 0$.

In the case of interest here, the matter fields consist of a pressureless dust fluid with $\rho_m \propto a^{-3}$, and a scalar field $\phi$. Unlike in the effective field approach reviewed in previous section, we assume the kinetic term of this scalar field to be of the standard canonical form and hence its Lagrangian is simply ${\cal L} = X - V(\phi)$. Correspondingly, for the scalar field the energy density and pressure are related by the standard equations
\begin{eqnarray}\label{rhoP_phi}
 \rho_\phi = \frac{1}{2}\d\phi^2 + V(\phi)~,~~P_\phi = \frac{1}{2}\d\phi^2 - V(\phi)~.
\end{eqnarray}

\subsection{Analytic Treatment}
\label{ss.hom-a}

As has been discussed in the Introduction, it is necessary to specifically choose the form of the potential for the scalar field $\phi$ in order to ensure that anisotropies never become dominant. For this reason, we choose the potential to have the ekpyrotic form of a negative-valued exponential function when the universe is in the contracting phase. One may simply take the same potential for the scalar that is given in \eqref{gV}. In the present subsection, however, we slightly deform the potential as follows,
\begin{eqnarray}\label{V_analytic}
 V(\phi) = \f{-V_o e^{-\sqrt{\frac{2}{p}} \frac{\phi}{M_{\rm Pl}}}} {\left(1 + \tf{3 p V_o}{4 \rho_c (1-3p)} \, e^{-\sqrt{\frac{2}{p}} \frac{\phi}{M_{\rm Pl}}}\right)^2} ~,
\end{eqnarray}
in order to simplify the analytic treatment \cite{Mielczarek:2008qw}. It is easy to check that the shape of the above potential is roughly the same as the potential \eqref{gV}.

The reason why we choose this specific potential is the following: when the only matter field is the scalar field $\phi$ with the potential \eqref{V_analytic}, there exists a particular solution whose contracting phase is precisely that of the ekpyrotic universe, namely a solution with a constant equation of state
\begin{eqnarray}\label{w_ekp}
 \om \equiv \frac{P_\phi}{\rho_\phi} = \frac{2}{3p} - 1 ~,
\end{eqnarray}
where $\om \gg 1$ for small $p$, and the scale factor is given by
\begin{eqnarray} \label{a-ekp}
 a(t) = \left(a_o M_{\rm Pl}^2 t^2 + 1 \right)^{p/2} ~,~~ a_o = \f{\rho_c}{3 p^2 M_{\rm Pl}^4},
\end{eqnarray}
where we have normalized the scale factor so that $a_B \equiv a(0) = 1$ at the bounce point $t_B=0$. Note that \eqref{a-ekp} is an attractor solution in the contracting phase and hence the trajectories of the scalar field $\phi$ approach the above result rapidly for a wide class of initial conditions. However, after the universe passes through the bounce and enters the expanding phase, typical solutions will deviate from the above form and quickly approach the solution where the space-time is dominated by a stiff fluid with $\om=1$. We will see this explicitly in the following subsection where we numerically study the dynamics of the system.

Finally, we take into account the pressureless dust field. It is easy to check that initially the background universe would have been dominated by the dust field before some transition time $t_e \ll -1/(\sqrt{a_o} M_{\rm Pl})$. The epoch of the matter-dominated contraction can be realized either by introducing a regular dust perfect fluid \cite{Cai:2012va} or a massive scalar field \cite{Cai:2013kja} (see also \cite{Cai:2008qw} based on the Lee-Wick construction).

Since the transition time $t_e$ from the matter-dominated period into the ekpyrotic contraction occurs far before the bounce, quantum gravity effects are negligible at this transition time and we can deal with the background system classically to an excellent approximation. Defining $a_e \equiv a(t_e)$ from Eq.\ \eqref{a-ekp}, the scale factor in the matter-dominated period is then given by
\begin{eqnarray}
 a(t) = a_e \left( \f{t - t_o}{t_e - t_o} \right)^{2/3} ~,~~
 t_o = t_e - \f{2}{3 H_e},
\end{eqnarray}
where $a_e$ and $H_e$ correspond to the values of the scale factor and the Hubble rate at the transition moment $t=t_e$ in order to ensure that both are continuous.

By combining the above results, we find that the universe starts in a matter-dominated phase and transits into a period of ekpyrotic contraction, and then bounces to an expanding branch with a fast-roll phase. Here we make the assumption that at the transition time (whether between matter-domination and ekpyrotic, or ekpyrotic and fast-roll), the scale factor and the Hubble rate evolve in a continuous fashion, while the equation-of-state parameter $\om$ is taken to be piece-wise constant. Under this approximation, we obtain the solution of the background universe with the following three phases corresponding to matter-dominated contraction, ekpyrotic contraction, and fast-roll expansion:
\begin{align}
 a(t) =
 \begin{cases}
  a_e \left( \f{t - t_o}{t_e - t_o} \right)^{2/3} & {\rm for} \: t \le t_e ~, \\
  \left(a_o M_{\rm Pl}^2 t^2 + 1 \right)^{p/2} & {\rm for} \: t_e \le t \le 0 ~, \\
  \left( \f{3 \rho_c}{M_{\rm Pl}^2} \, t^2 + 1 \right)^{1/6} & {\rm for} \: t \ge 0 ~.
 \end{cases}
\end{align}
Note that after the bounce, the equation-of-state will rapidly evolve to $\om=1$ and the universe will enter a period of fast-roll expansion. One of the main goals of this paper is to track cosmological perturbations from the contracting branch to the expanding one, and we will see, both from analytic arguments and numerical evidence, that in the matter-ekpyrotic model the transition to fast-roll at the bounce point does not affect the long-wavelength modes that are relevant to cosmological observations.  Therefore, when studying perturbations in this context, the period of fast-roll expansion can safely be ignored.

Before numerically studying the effective dynamics of the matter-ekpyrotic universe in LQC, it is helpful to collect some equations that will be useful for the study of the dynamics of cosmological perturbations.  Here we recall some equations that only hold in the classical limit, when $t \ll -M_{\rm Pl}^{-1}$.  First, in the ekpyrotic era, the scale factor and the Hubble rate are given by
\be \label{class-t}
a(t) = \left(\sqrt{a_o} M_{\rm Pl} t \right)^p, \quad
H(t) = \f{p}{t},
\ee
and therefore, since $t_e \ll -M_{\rm Pl}^{-1}$, the Hubble rate at the transition time between matter-domination and the ekpyrotic phase is $H_e = p / t_e$.  It is convenient to work in conformal time $\eta$ in order to study cosmological perturbations, defined by $\dd \eta = \dd t / a(t)$, which is easily solved for the scale factor \eqref{class-t} giving
\be
(- M_{\rm Pl} t) = a_o^{\t{p}/2} \left[ -(1-p) M_{\rm Pl} \, \eta \right]^{\tf{1}{1-p}}~, ~~
\t{p} = \f{p}{1-p}.
\ee
From this relation, it is straightforward to express the scale factor in terms of the conformal time,
\be
 a(\eta) = \Big[ \sqrt{a_o} (1-p) M_{\rm Pl} (-\eta) \Big]^{\t{p}}~,
\ee
which holds for $\eta_e \le \eta \ll M_{\rm Pl}^{-1}$, where $\eta_e$  is the conformal time corresponding to $t_e$.  It follows that the conformal Hubble rate $\mH = a'/a$ (where the prime denotes differentiation with respect to the conformal time $\eta$) during the classical portion of the ekpyrotic contraction is given by
\be
 \mH(\eta) = \f{p}{(1-p) \, \eta} ~,
\ee
and therefore the conformal Hubble rate at the transition time is $\mH_e \equiv \mH(\eta_e) = p/[(1-p) \eta_e]$.  An equation that will be useful later is the relation between the conformal Hubble rate and the proper time Hubble rate at the transition time,
\be \label{hubbles}
 H_e = \f{\mH_e^{1+\t{p}}}{\left( \sqrt{a_o} p M_{\rm Pl} \right)^{\t{p}}} ~.
\ee

Finally, the scale factor as a function of conformal time during the initial matter-dominated phase ($\eta \le \eta_e$) is
\be \label{scale-eta}
 a(\eta) =  a_e \left( \f{\eta - \eta_o}{\eta_e - \eta_o} \right)^2~,
\ee
where $\eta_o = \eta_e - 2/\mH_e$, and the conformal Hubble rate during the matter-dominated contraction is
\be \label{conf-H}
\mH(\eta) = \f{2}{\eta - \eta_o}~.
\ee

It is important to keep in mind that the Eqs.~\eqref{class-t}--\eqref{conf-H} only hold in the classical limit, namely where $t \ll - M_{\rm Pl}^{-1}$ and $\eta \ll - M_{\rm Pl}^{-1}$.

\subsection{Numerics of the Background Dynamics}
\label{ss.hom-n}

In order to explicitly show that a non-singular bounce occurs in LQC without violating any energy conditions, we numerically solve the background evolution. In the numerical computation the matter content consists of a pressureless ($P_m=0$) dust fluid and a canonical scalar field with the potential \eqref{gV}. Note that this potential is slightly different from the potential \eqref{V_analytic} used in the analytic treatment, but their shapes are the same and in both cases the slopes of these potentials are steep enough to drive the same ekpyrotic contracting phase that was studied in detail in \cite{Cai:2012va}. For our numerical studies, we choose the potential \eqref{gV} in order to better be able to compare the results obtained in the effective field model of the matter-ekpyrotic bounce. A key observation is that the dynamics of the canonical scalar field in LQC is very similar to that of the non-conventional scalar in the effective field model proposed in \cite{Cai:2012va}. This similarity nicely demonstrates that the physics of LQC can be mimicked by effective theory models even though the bounce occurs in the deep Planck regime.

\begin{figure*}[th]
\centering
\subfloat[]
{\includegraphics[width=0.3\textwidth]{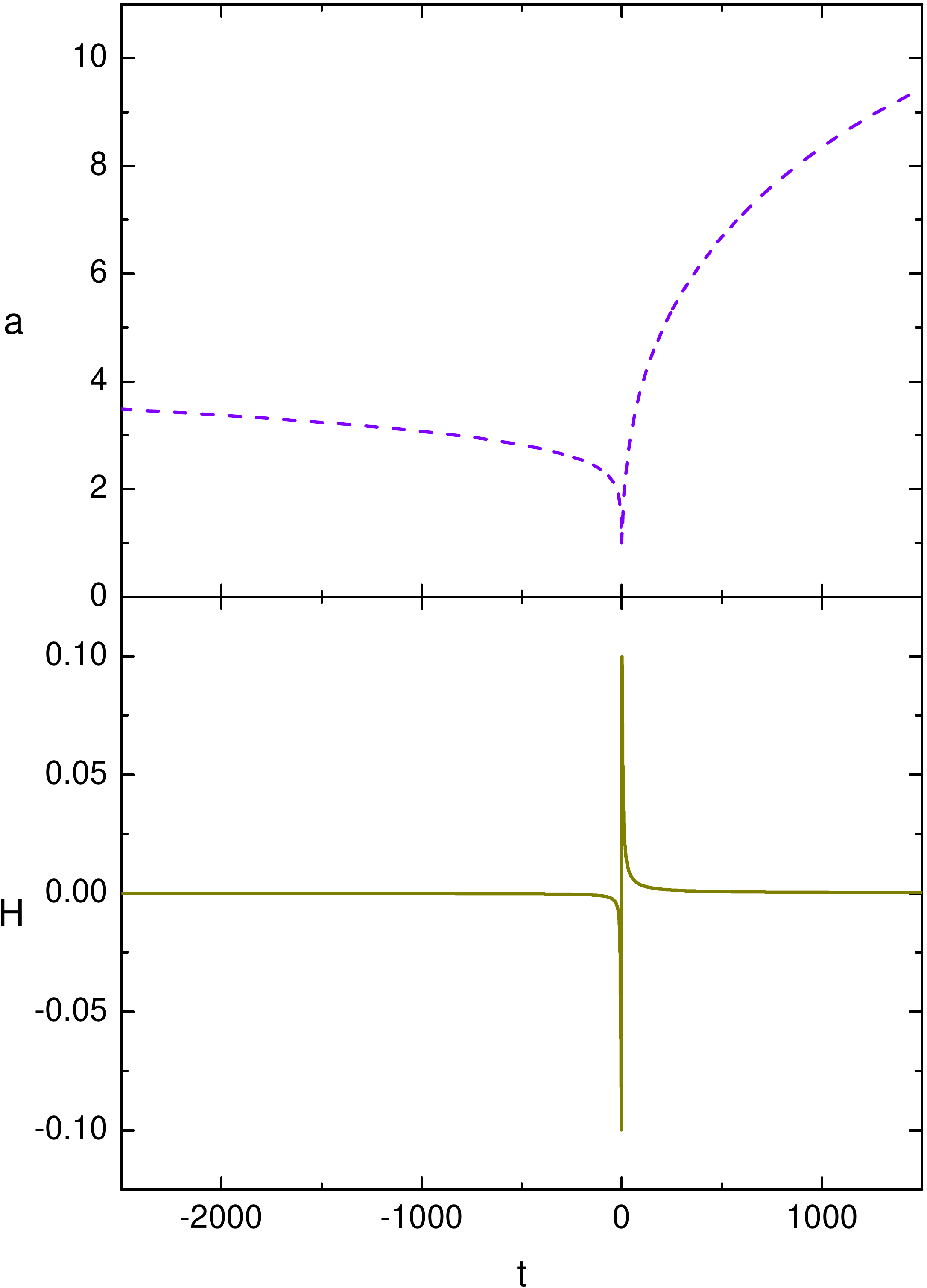} \label{f1a}}
 \qquad
\subfloat[]
{\includegraphics[width=0.3\textwidth]{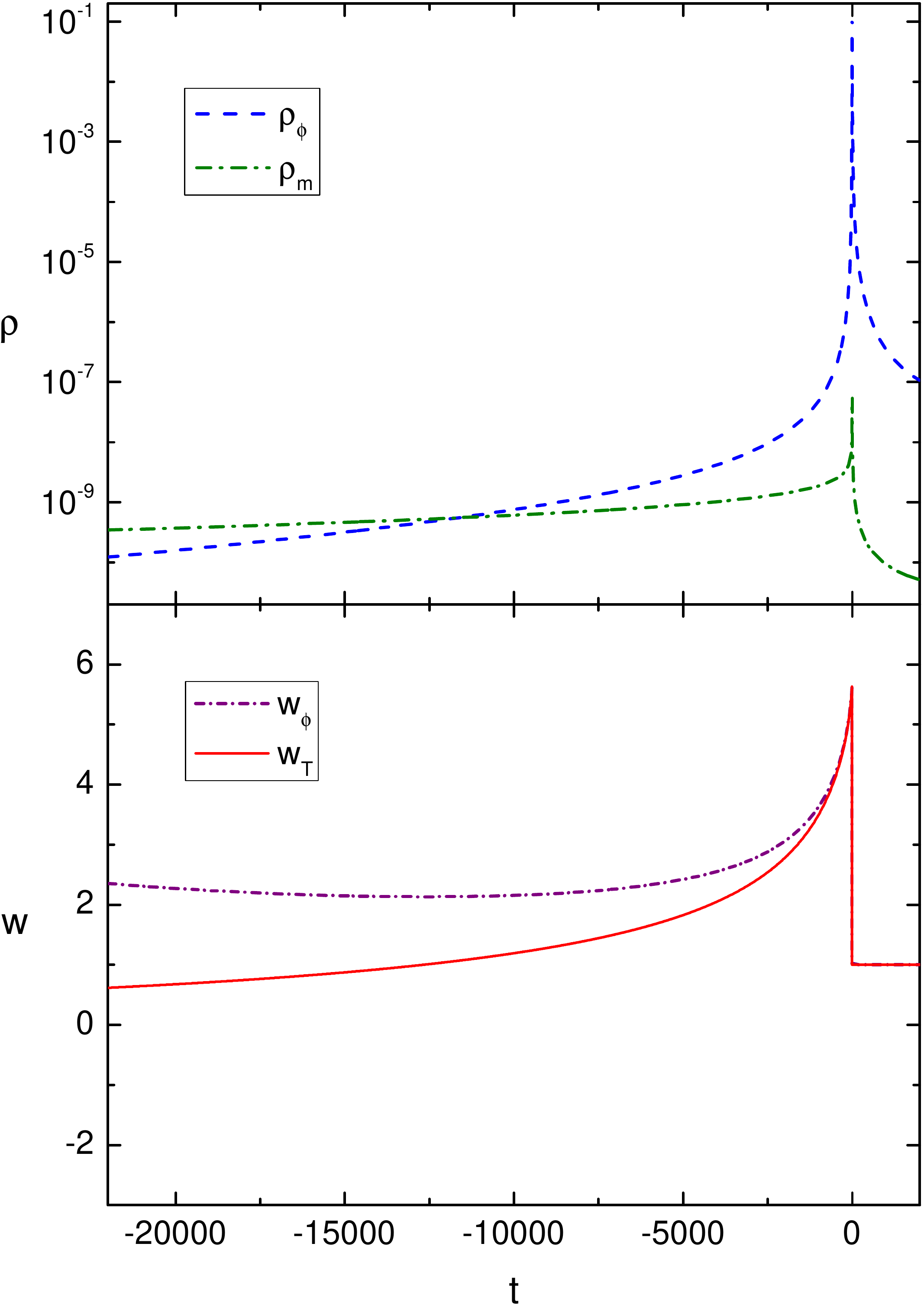} \label{f1b}}
 \qquad
\subfloat[]
{\includegraphics[width=0.3\textwidth]{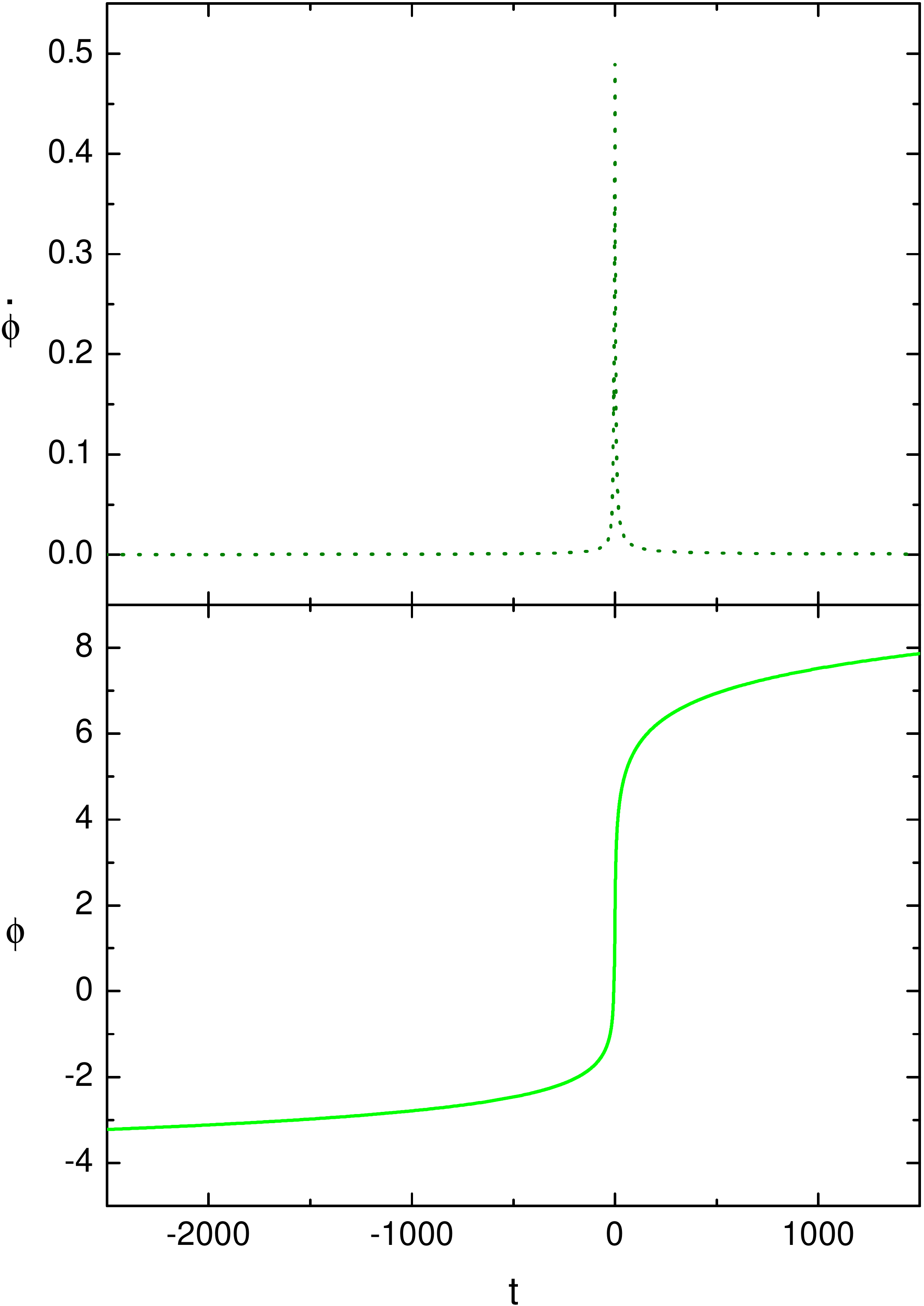} \label{f1c}}
\caption[]{\footnotesize \hangindent=10pt
Plot of the numerical evolution of the matter-ekpyrotic universe in LQC. The horizontal axis denotes the cosmic time $t$. The left panel shows the evolution of the scale factor $a$ (the violet dashed curve) and the Hubble parameter $H$ (the dark yellow solid curve). The middle panel plots the evolution of the energy density of the scalar field $\rho_\phi$ (the blue dashed curve) and that of dust fluid $\rho_m$ (the green dash-dot curve) as well as the equation of state parameter of the scalar field $\omega_\phi$ (the purple short dash-dot curve) and the effective ``total'' equation of state $\omega_T$ (the red solid curve). The right panel gives the dynamics of the scalar field $\phi$ (the green solid curve) and its time derivative $\dot\phi$ (the olive dotted curve). The background parameters chosen for the numerics are given in Eq.~\eqref{para_bg}.}
\label{f1}
\end{figure*}

We work in units of the reduced Planck mass $M_{\rm Pl}$ for all dimensional parameters. Specifically, for the numerical simulations we choose
\begin{align}\label{para_bg}
 V_0 = 8 \times 10^{-3} ~,~~ p = 0.1 ~,~~ b_V = 5 ~,~~ \rho_c = 0.12 ~,
\end{align}
and take the initial value of the energy density of dust fluid as $\rho_{m}^i = 2.1 \times 10^{-10}$. The corresponding results are given in Fig.~\ref{f1}.

From Fig.~\ref{f1a}, we see explicitly that the Hubble parameter (as depicted by the dark yellow solid line in the lower panel) remains finite as it evolves from a negative to a positive value, showing that the universe transits from the contracting phase to the expanding phase in a non-singular fashion. The same conclusion is also reached by looking at the evolution of the scale factor (as depicted by the violet dashed line in the upper panel), which decreases before reaching a minimal non-zero value and then increasing monotonically thereafter. The bounce point occurs at the moment $t_B=0$ where the scale factor shrinks to the minimal value $a_B=1$.

Fig.~\ref{f1b} shows the evolution of the energy densities and the equation-of-state parameters in the model. From the upper panel, one sees that initially the universe was dominated by the pressureless dust fluid (the green dash-dot line). As the space-time contracts, the contribution of the scalar field (the blue dashed line) becomes dominant and initiates a period of ekpyrotic contraction. After the bounce the universe evolves into a fast-roll expanding phase with an effective equation of state given by $\omega = 1$ as depicted by the lower panel of Fig.~\ref{f1b}.

We also numerically plot the dynamics of the background scalar $\phi$ (by the green solid line) and its time derivative $\dot\phi$ (by the olive dot line) in Fig.~\ref{f1c}. One can see that $\phi$ evolves monotonically from negative infinity to positive infinity. Around the bounce point, $\dot\phi$ reaches its maximal value and then falls back to a small value.

The background dynamics of the matter-ekpyrotic universe in LQC closely resemble what was found in the effective field model with a non-standard kinetic term \cite{Cai:2012va}. Indeed, the evolution of the scale factor, the Hubble rates, the effective equations of state and even the scalar fields themselves are extremely similar in the two settings. We will show in the next section that the dynamics of cosmological perturbations also match.

\section{Scalar Perturbations}
\label{s.scal}

After having studied the background dynamics, the next step is to determine the evolution of the primordial cosmological perturbations. We shall focus on scalar perturbations as from an observational point of view they are the most relevant.  We also restrict our analysis to linear perturbations (note that the contribution of non-linear fluctuations in the contracting phase is very small, as analyzed in \cite{Cai:2009fn}), in which the Fourier modes evolve independently and therefore it is possible to track the evolution of each mode separately. For a comprehensive review of cosmological perturbation theory in general relativity, see \cite{Mukhanov:1990me}.

In LQC, there are quantum gravity effects coming from expressing the Hamiltonian constraint in terms of holonomies of the Ashtekar-Barbero connection that become significant when the space-time curvature nears the Planck scale. Effective equations of motion that capture the main quantum gravity effects of LQC in the dynamics of cosmological perturbations can be derived either from the Hamiltonian constraint of lattice LQC \cite{WilsonEwing:2011es} or by following the anomaly freedom algorithm \cite{Cailleteau:2011kr}. (A hybrid quantization approach yields similar, although slightly different results \cite{FernandezMendez:2012vi}.) Since the quantum gravity effects are negligible when the universe is far from the bouncing phase, the effective equations for cosmological perturbations rapidly approach the standard Mukhanov-Sasaki equations of general relativity in the small curvature limit, both in the contracting and expanding epochs. However, quantum gravity effects do become important in the vicinity of the bounce.  In addition, as the singularity is resolved in the homogeneous background space-time, the perturbations pass through the bounce smoothly and it is possible to explicitly determine their evolution from the distant past to the post-bounce expanding phase. It is interesting to compare the results obtained in LQC to those found in effective theory models; as we shall see they are very similar.

Note that we will not attempt to address the trans-Planckian problem, which is an important conceptual issue for inflationary cosmology \cite{Brandenberger:1999sw}. The effective equations that we shall use here come from lattice LQC which can only describe perturbations with a wavelength greater than the Planck length \cite{WilsonEwing:2011es}. Although trans-Planckian modes can in fact be handled in the hybrid quantizations of cosmological perturbations \cite{FernandezMendez:2012vi}, at this point in time it is not clear whether trans-Planckian modes can exist or not in full loop quantum gravity and therefore we will ignore this issue here.  In any case, the modes that would have been trans-Planckian during the bounce in the matter-ekpyrotic model are not observationally relevant and therefore from a purely practical point of view it is reasonable to ignore these modes.

Also, in this paper we focus on the holonomy corrections of LQC that are responsible for the bounce; for information regarding inverse triad effects, see e.g.\ \cite{Bojowald:2008jv}.

To study scalar cosmological perturbations, the metric and the matter fields are perturbed to linear order. As the fluctuations of the matter and geometry degrees of freedom are dynamically related, there is only one degree of freedom and it is convenient to introduce the gauge-invariant quantity related to the curvature perturbation in the uniform field gauge by $v=z\zeta$, which is the famous Mukhanov-Sasaki variable \cite{Sasaki:1986hm}.

In terms of the Fourier modes of the variable $v$, the effective perturbation equation in LQC can be written as
\begin{eqnarray} \label{ms-scal}
 v_k'' + \left( c_s^2 k^2 - \frac{z''}{z} \right) v_k = 0~,
\end{eqnarray}
where the prime denotes the derivative with respect to the conformal time $\eta$ and the subscript $k$ denotes the comoving wave number of the corresponding Fourier mode. The above perturbation equation involves two parameters with one being the square of the sound speed
\begin{eqnarray}
 c_s^2 = 1-\frac{2\rho}{\rho_c}~,
\end{eqnarray}
and the other is related to the background evolution via
\begin{eqnarray}
 z = a \frac{\sqrt{\rho+P}}{H} ~ = ~ a \, M_{\rm Pl} \sqrt{\frac{3(1+\om)}{1-\rho/\rho_c}}~,
\end{eqnarray}
where in the last equality $\om$ may in general depend on time.  However, in this section we shall make the approximation of a constant $\om = 0$ in the matter-dominated epoch, and a constant $\om = 2/3p - 1$ in the ekpyrotic phase.

The sound speed plays the role of governing the propagation speed of the fluctuations. In LQC, the bouncing phase is triggered when the background energy density $\rho$ approaches the critical density $\rho_c \sim M_{\rm Pl}^4$. In particular, when $\rho$ evolves into the regime of $\rho>\rho_c/2$, we get $c_s^2<0$. This is the precise feature that was previously observed in the effective field model of the matter-ekpyrotic bounce cosmology discussed in Sec \ref{s.eff}. The appearance of a negative sound speed squared, from the viewpoint of effective field theory, indicates a dangerous gradient instability in the ultraviolet scales; from the perspective of LQC, however, it is an effect due to quantum gravity. This is an indication that the effective field approach developed in \cite{Cai:2012va} is reliable even during the bounce.

Also, since from an observational point of view the relevant variable is the gauge-invariant curvature perturbation
\be
\zeta_k = v_k / z,
\ee
it is convenient to write down the equation of motion for $\zeta_k$ in LQC. For our purposes, it will be sufficient to work in the long wavelength limit of modes that are outside the Hubble radius, in which case the equation of motion for $\zeta_k$ simplifies to
\be \label{zeta}
 \zeta_k'' + \f{2 z'}{z} \zeta_k' = 0 ~.
\ee

\subsection{Analytic Treatment}
\label{ss.scal-a}

We shall start during the matter-dominated phase, when the effects of LQC are completely negligible and the perturbation equation rapidly approaches the standard equation of general relativity for $v_k$,
\be
 v_k'' + k^2 v_k - \f{2}{(\eta-\eta_o)^2} v_k = 0 ~.
\ee
The solution to this differential equation is the Hankel function
\begin{align}
 v_k(\eta) = \sqrt{-(\eta-\eta_o)} & \Big[ A_1 H_{3/2}^{(1)}(-k (\eta - \eta_o)) \nonumber\\
 & + A_2 H_{3/2}^{(2)}(-k (\eta - \eta_o)) \Big] ~,
\end{align}
and setting the initial conditions at $\eta\to-\infty$ to be the quantum vacuum gives
\be
 v_k(\eta) = \sqrt \f{-\pi (\eta-\eta_o)}{4} H_{3/2}^{(1)}[-k (\eta - \eta_o)] ~.
\ee
For modes that exit the Hubble radius during the matter-dominated epoch, it is possible to use the small argument expansion of the Hankel function, and then $v_k(\eta)$ takes the form
\be \label{v-matt}
 v_k(\eta) = \f{\sqrt\pi}{\sqrt{2} \, \Gamma\!\left(\tf{5}{2}\right)} \cdot \f{k^{3/2}}{\mH(\eta)^2}
 + i \, \f{\Gamma\!\left(\tf{3}{2}\right)}{\sqrt{2 \pi}} \cdot \f{\mH(\eta)}{k^{3/2}} ~,
\ee
where the time dependence is expressed in terms of the conformal Hubble rate that is given in \eqref{conf-H}. Note that the second term is scale-invariant.

Now let us move on to the ekpyrotic phase. In this setting, the perturbation equation becomes (again in the classical regime $\eta \ll -M_{\rm Pl}^{-1}$)
\be
 v_k'' + k^2 v_k - \f{p(1-p)}{(p-1)^2 \, \eta^2} v_k = 0 ~,
\ee
and the solution is
\be
 v_k(\eta) = B_1 \sqrt{-\eta} J_n(-k \eta) + B_2 \sqrt{-\eta} Y_n(-k \eta) ~,
\ee
with
\be
 n = \f{1-3p}{2(1-p)} \approx \f{1}{2} - p ~,
\ee
where the second relation holds approximately for small $p$, which is the case of interest during the ekpyrotic phase.

For modes outside the Hubble radius,
\be \label{v-ekp1}
 v_k(\eta) = \t{B_1} (-\eta)^{n+1/2} + \t{B_2} (-\eta)^{-n+1/2} ~,
\ee
where we have absorbed some overall constants into $\t{B_1}$ and $\t{B_2}$. The prefactors $\t{B_1}$ and $\t{B_2}$ can be determined by matching with the solution \eqref{v-matt} from the matter-dominated epoch by imposing that $v_k$ and $v_k'$
are continuous at the transition time $\eta_e$.

As the quantity of interest from an observational point of view is the comoving curvature perturbation $\zeta_k = v_k/z,$ where $z \sim (-\eta)^{p/(1-p)}$ in the ekpyrotic epoch (here again we ignore for now the very small time-frame where LQC effects are important), we are only interested in the terms in $\zeta_k$ that do not vanish as the bounce is approached. A quick inspection shows that the term with the $\t{B_1}$ factor goes to zero as $\eta \to 0$, while the $\t{B_2}$ term in $\zeta_k$ is constant in time.  Therefore, the only term that is observationally relevant is the second one in \eqref{v-ekp1}, and so it is only necessary to determine $\t{B_2}$. Imposing continuity in $v_k$ and $v_k'$ at $\eta=\eta_e$ gives
\be
 \t{B_2} = i \, \f{\Gamma\!\left(\tf{3}{2}\right)}{\sqrt{2 \pi}} \cdot f(p) \cdot \mH_e^{1+\t{p}} \cdot k^{-3/2} + \alpha k^{3/2} ~,
\ee
where $f(p) = 1 + O(p \log p)$ is a function that is close to 1 for small $p$. The precise value of $\alpha$ is not important here because for the long wavelength modes that are relevant from an observational point of view, $k^{-3/2}$ dominates $k^{3/2}$ and therefore the second term here is negligible and will be dropped from here on.

A key result here is that the amplitude of the scale-invariant term in the super-Hubble modes is related to the conformal Hubble rate $\mH_e$ at the transition point between the matter-dominated and the ekpyrotic phases of the space-time. Therefore, the greater the Hubble rate is at this transition time, the greater the amplitude of the perturbations will be and vice versa.

Therefore, before the bounce the dominant contribution to the scalar perturbations has a constant amplitude in time given by
\be \label{scal-sol}
 \zeta_k(\eta) = \f{\sqrt{p} ~ \t{B_2}}{\sqrt{2} \, M_{\rm Pl}^{1+\t{p}} [\sqrt{a_o} (1-p)]^{\t{p}}} ~.
\ee
Note that we have dropped the term which decays as $\eta$ approaches zero, while the dominant term in $\t{B_2}$ goes as $k^{-3/2}$ and so is scale-invariant.

This is the classical solution in the ekpyrotic phase for modes that exited the Hubble radius during the matter-dominated epoch.  Now it is necessary to evolve this solution through the quantum gravity regime where the bounce occurs in order to determine the form of the scalar perturbations after the bounce. In this case, this evolution is trivial as $\zeta_k$ already satisfies the differential equation \eqref{zeta}. Therefore, after the bounce, $\zeta_k$ still has the exact same form given in \eqref{scal-sol}.

Note that in general the evolution across the bounce is non-trivial: if the dominant mode in the scalar perturbations is not constant with respect to time, then there will be LQC effects that will appear in the differential equation from the $z'/z$ term which will affect the final power spectrum.  However, in this particular case the dominant term is time-independent, and so the evolution across the bounce is trivial.

Therefore, the power spectrum for scalar perturbations in this scenario is
\be\label{Pz}
 P_\zeta(k) = \f{k^3}{2 \pi^2} |\zeta_k|^2 = {\f{p}{16\pi^2}} \cdot \t{f}(p) \cdot \frac{H_e^2}{M_{\rm Pl}^2} ~,
\ee
where $\t{f}(p) = 1 + O(p \log p)$, and we used the relation \eqref{hubbles} to replace the occurrences of $\mH_e$ by $H_e$.

From \eqref{Pz} we see that the amplitude of the power spectrum of the primordial curvature perturbations is determined by the Hubble parameter at the end of the era of matter-dominated contraction. This is because the curvature perturbation, which is originally increasing on super-Hubble scales in the matter-dominated contracting phase, freezes after the universe enters into the period of ekpyrotic contraction. Quantum gravity effects play a limited role (beyond the key resolution of the big-crunch/big-bang singularity in the background homogeneous space-time) as they do not affect the propagation of long wavelength modes of constant amplitude.

Note that this result is significantly different from what occurs in the matter bounce scenario (without an ekpyrotic scalar field) in LQC where it was observed that the LQC parameter $\rho_c$ has to be taken to be very small in order to reproduce the observed $P_\zeta(k) \sim O(10^{-9})$ \cite{Spergel:2003cb, Ade:2013zuv}.  This is because the strength of the space-time curvature at the bounce ---which is proportional to $\rho_c$--- determines the amplitude of the scale-invariant perturbations after the bounce \cite{WilsonEwing:2012pu}. As pointed out above, this is avoided in the matter-ekpyrotic bounce scenario as the super-Hubble modes stop growing when the universe enters the ekpyrotic phase and so their amplitude is determined by the energy scale of the transition from matter-domination to the ekpyrotic phase, no matter the value of $\rho_c$.  Also, note that if there is no dust field, then it is possible to generate a scale-invariant power spectrum via the entropic mechanism if there are two ekpyrotic scalar fields, in which case the resulting amplitude depends on the energy scale at the moment of transferring the entropy modes into adiabatic ones and also on the specific potentials of the ekpyrotic scalar fields \cite{Wilson-Ewing:2013bla, ekpyrotic}.

An important point is that this result is in agreement with the conclusion based on the effective field approach studied in \cite{Cai:2012va}. Therefore, the comparison of the perturbation analyses in these two approaches again demonstrates the reliability of the effective field treatment for non-singular bounce models in cosmology.

\subsection{Numerical Analysis of the Perturbations}

\begin{figure*}[th]
\centering
\subfloat[][]
{\includegraphics[width=0.3\textwidth]{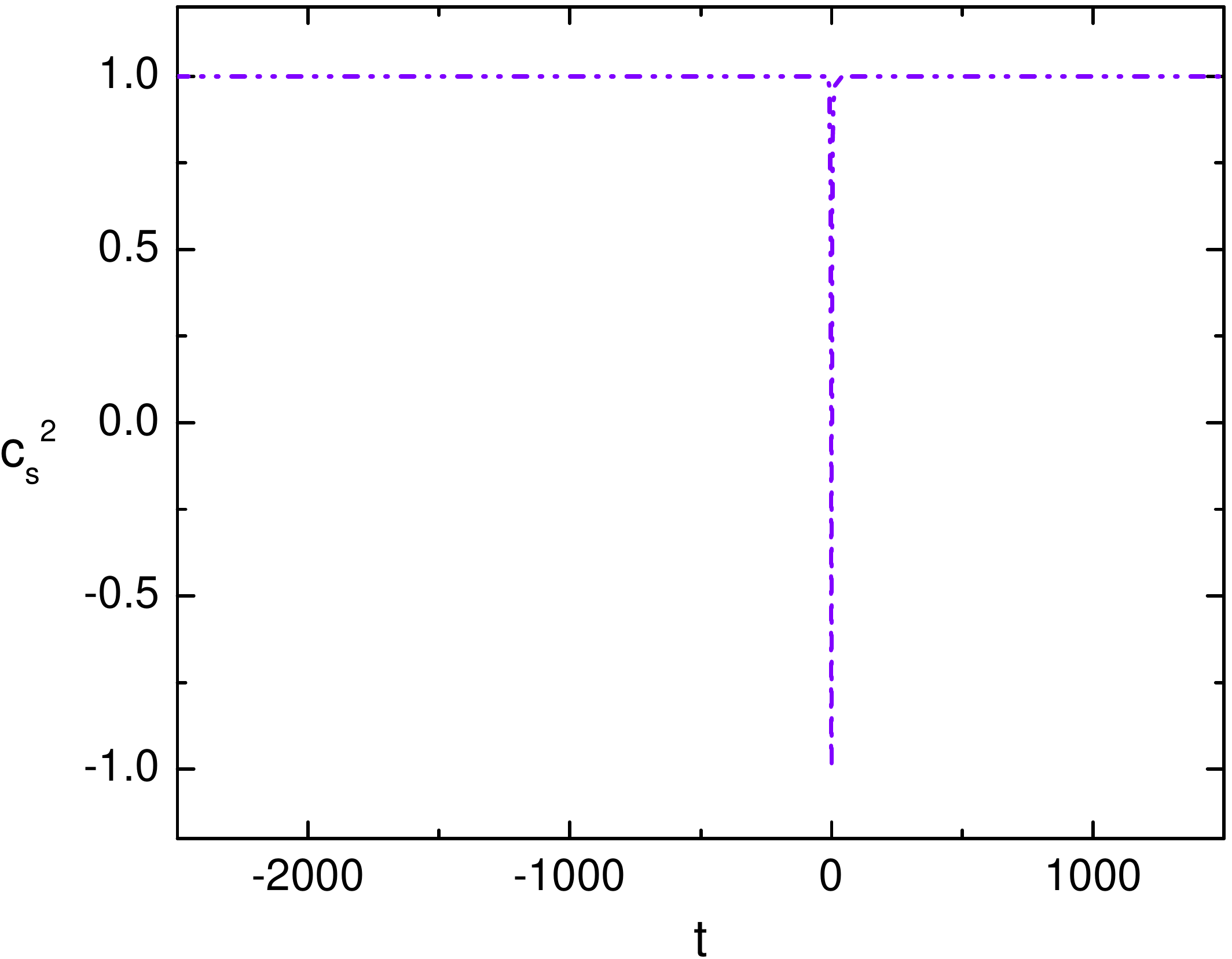} \label{f2a}}
 \qquad
\subfloat[][]
{\includegraphics[width=0.3\textwidth]{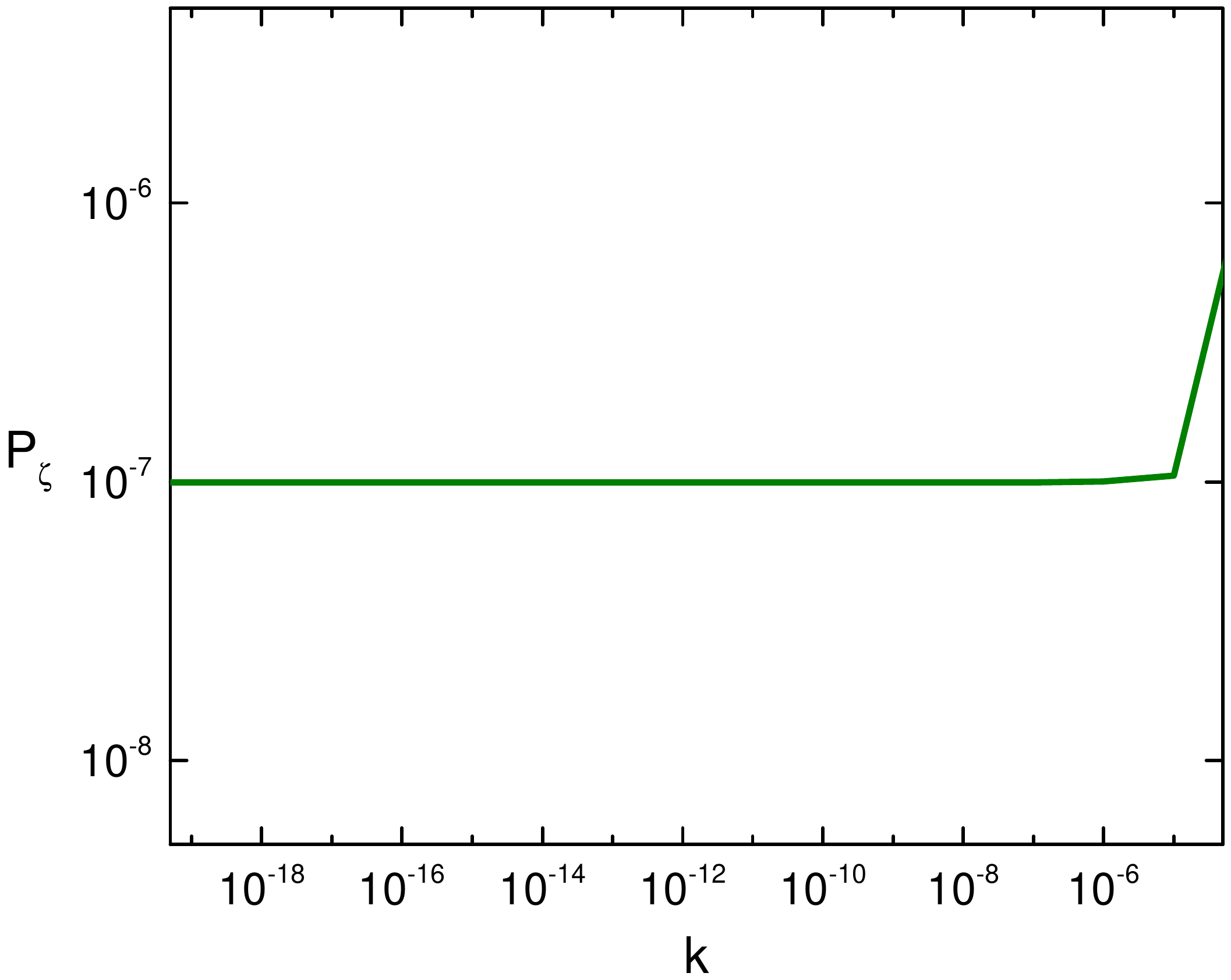} \label{f2b}}
\caption{\footnotesize \hangindent=10pt
Numerical plot of the dynamics of cosmological perturbations in the matter-ekpyrotic model in LQC. The left panel shows the evolution of the sound speed squared $c_s^2$ (the violet dash-dot-dot curve) as a function of cosmic time. The right panel depicts the primordial power spectrum $P_\zeta$ (the green solid curve) as a function of the comoving wave number $k$. The initial conditions for the background field and model parameters are the same as for Fig.~\ref{f1} and are given in \eqref{para_bg}. The initial conditions for the curvature perturbations are that they are initially quantum vacuum fluctuations in the early matter-dominated contracting phase.
}
\label{f2}
\end{figure*}

Following what was done for the background analysis, in the present subsection we numerically study the dynamics of the primordial perturbations for the matter-ekpyrotic model in LQC. Specifically, the initial values of the background parameters are those given in \eqref{para_bg}, and the primordial perturbations are taken to initially be quantum vacuum fluctuations in the early matter-dominated contracting phase. The numerical results are shown in Fig.~\ref{f2}.

The evolution of the sound speed squared $c_s^2$, which characterizes the gradient instability of the perturbation modes, is shown in Fig.~\ref{f2a}. For the vast majority of the evolution it is equal to unity and hence the propagation of the primordial fluctuations are safe against the gradient instability. When the universe enters the bouncing phase, however, this parameter becomes negative and falls to its minimal value of $-1$ at the bounce point. A similar phenomenon was also observed in the effective field approach of bounce cosmology studied in \cite{Cai:2012va}. From the viewpoint of effective field theory, the sign change in $c_s^2$ implies that there exists a certain gradient instability for the perturbation modes at ultraviolet scales during the bouncing phase. As the sign of the sound speed squared also changes around the bounce point in LQC, we see that the effective theory actually captures this property surprisingly well.

In Fig.~\ref{f2b} we show the scale dependence of the primordial power spectrum $P_\zeta$ after the bounce as a function of the comoving wave number $k$. The figure is plotted in a log-log scale for a better visual effect. We can interestingly observe that the power spectrum is nearly scale-invariant for the perturbation modes in the infrared regime but has a blue tilt for the ultraviolet modes. The scale dependence changes around the comoving wave number $k\sim 10^{-5}$, beyond which point the perturbation modes exit the Hubble radius during the ekpyrotic phase. Thus, this result explicitly shows that the primordial power spectrum of curvature perturbations is scale-invariant only for the modes which exit the Hubble radius in the matter-dominated contracting phase.

Note that in our numerical computation we chose a specific class of model parameters to produce the best visual effect and hence the energy scale of the transition from the matter-dominated contraction to the ekpyrotic phase is slightly higher than the GUT scale. As a result, the amplitude of the power spectrum in the case under consideration is of the order of $O(10^{-7})$ which is higher than the observed amplitude in the CMB of $O(10^{-9})$. It is easy to modify the amplitude and be in agreement with observations simply by changing the initial conditions to a smaller initial value of $\rho_m$ so that the transition to the ekpyrotic era occurs earlier and at a lower energy scale, hence making $H_e$ smaller. The key point here, however, is that the modes that exit the Hubble radius during the matter-dominated era have a scale-invariant spectrum.

\subsection{Tensor Perturbations}
\label{s.tens}

In addition to the scalar perturbations studied in the two previous subsections, primordial tensor perturbations are also generated in the contracting branch of the universe. The latest observations provide an upper bound to the tensor-to-scalar ratio of $r < 0.11$ ($95\%$ CL) \cite{Ade:2013uln}. As is well known, when the universe has a constant equation of state and is far away from the bouncing phase, the equation of motion for the tensor fluctuations is analogous to that of the scalar curvature perturbations. Accordingly, assuming that the scalar and tensor perturbations are initially in the quantum vacuum state, they grow at the same rate in a matter-dominated phase of contraction. As a result, the tensor-to-scalar ratio in the traditional scenario of the matter bounce is typically too large to be consistent with the observational data \cite{Cai:2008qw} (although not in LQC \cite{WilsonEwing:2012pu}).

There are three main approaches that can resolve this issue. The first is to take into account the entropy modes by involving other matter fields, such as the curvaton mechanism \cite{curvaton} considered in the matter bounce in \cite{Cai:2011zx}. In this mechanism the conversion from entropy perturbations to curvature perturbations occurs during the bouncing phase and can efficiently enhance the amplitude of primordial adiabatic fluctuations without affecting the tensor fluctuations. A closely related approach is to have a purely ekpyrotic contracting phase and generate scale-invariant curvature perturbations via the entropic mechanism, in which case the tensor perturbations have a vanishingly small amplitude; this was studied within the context of LQC in \cite{Wilson-Ewing:2013bla}. The second method is to achieve a gravitational amplification effect on the curvature perturbations from the gradient instability of $c_s^2$ changing sign during the bounce. This enhancement effect can be sufficiently large if the bouncing phase lasts for a very short period while the bounce energy scale is low enough \cite{Cai:2013kja}. However, in the specific model considered in the present paper, we find that the enhancement from the gravitational amplification is of order unity as the bounce occurs at a high energy scale. The third method of lowering the tensor-to-scalar ratio in LQC is in the pure matter bounce scenario without an ekpyrotic field.  Further, if the dust fluid has a small negative non-vanishing pressure $P = -\epsilon \rho$, this gives a small tensor-to-scalar ratio and also a slight red tilt to the power spectrum of the curvature perturbations \cite{WilsonEwing:2012pu}. Note however that in this last case, anisotropies may become important around the bounce point.

In the matter-ekpyrotic scenario in LQC, the tensor-to-scalar ratio will typically be larger than the observational bound, and therefore it will be necessary to find a mechanism to damp $r$, possibly in one of the ways described here.  In the present work, however, our focus is on the comparison between the LQC description and the effective field approach of non-singular bouncing cosmologies, and hence we leave this particular topic for future study.

\section{Conclusions}
\label{s.disc}

In LQC, the big-bang and big-crunch singularities are generically resolved without any need to introduce the non-standard kinetic operators that are required in the effective field approach. In this paper we study the cosmological dynamics of a canonical scalar field with an ekpyrotic-like potential and a dust field in the context of LQC. With this choice of matter content, the universe initially starts its evolution with an epoch of matter-dominated contraction, then it transits into a period of ekpyrotic contraction, and afterwards experiences a smooth bouncing phase which connects to a regular thermal expansion. The background evolution obtained based on quantum gravity effects in LQC is very similar to that described by the effective field approach involving non-canonical kinetic operators.

One promising feature of the matter-ekpyrotic bounce cosmology is that it generates a scale-invariant power spectrum of primordial curvature perturbations. The scale-invariant perturbations are generated in the matter-dominated contracting phase and their amplitude is determined by the Hubble rate at the end of this phase, which in our model is denoted by $H_e$. We show that this scale invariance is well preserved across the bounce and compare this with the results found in the effective field approach (for example see the study of \cite{Finelli:2001sr} based on matching conditions \cite{Hwang:1991an, Deruelle:1995kd} and \cite{Cai:2007zv, Cai:2008qw, Xue:2013bva} using various field theory models). Interestingly, the effective theory model captures much of the relevant physics of LQC even though the effects of LQC only become important in the deep Planck regime.

This is somewhat surprising as effective field theories are typically only expected to be reliable when quantum gravity effects are small, but here we have seen that the main properties of the evolution through the bounce in LQC ---of both the homogeneous background and the perturbations--- are captured in the effective theory models, even though these effects occur in the deep non-perturbative regime of LQC.  In the effective theory approach, the bounce usually occurs well below the Planck regime in order to ensure that quantum gravity effects will not ruin the validity of the effective theory.  However, we have seen here that effective theory models in fact capture the key qualitative features of LQC, and therefore can act as toy models for full quantum gravity when the bounce energy scale of these models is in the Planck regime. This suggests that effective field theory models in cosmology may be more reliable at high energy scales than initially expected.

\acknowledgments
We would like to thank Ivan Agull\'o, Robert Brandenberger and Luca Bombelli for helpful discussions. The work of YFC is supported in part by an NSERC Discovery grant and by funds from the Canada Research Chair program. This work is supported in part by a grant from the John Templeton Foundation. The opinions expressed in this publication are those of the authors and do not necessarily reflect the views of the John Templeton Foundation.

\end{document}